\documentclass[epj]{svjour}
\usepackage{graphicx}
\usepackage{amssymb}
\usepackage{dcolumn}
\usepackage{amsmath}
\usepackage{bm}
\usepackage{textcomp}
\usepackage{epstopdf}
\usepackage{color,url}
\usepackage{ulem}
\usepackage[toc,page,title,titletoc,header]{appendix}
\usepackage[colorlinks,
            linkcolor=blue,
            anchorcolor=blue,
            citecolor=blue,
            urlcolor=blue
            ]{hyperref}
\usepackage{txfonts}
\usepackage{caption}
\usepackage{lineno}

\begin{document}

\title{The equilibrium and dynamical cumulants of QCD chiral order parameter with parametric Landau free energy}
\author{Lijia Jiang\inst{1} \and Horst St\"ocker\inst{2} \and Jun-Hui Zheng\inst{1}
\thanks{{e-mail:} \href{mailto:junhui.zheng@nwu.edu.cn}{junhui.zheng@nwu.edu.cn} (corresponding author).}%
 }

\institute{Institute of Modern Physics, Northwest University, 710127 Xi'an, China \and Frankfurt Institute for Advanced Studies, 60438 Frankfurt am Main, Germany}

\date{Received: date / Revised version: date}

\abstract{By linearly parameterizing the QCD Landau free energy near the critical point in the baryon chemical potential and temperature plane, we study the fluctuations of the QCD chiral order parameter field (the $\sigma$ field) in the equilibrium case and dynamical phase transition, respectively. By setting the system size to the typical size of the QGP fireball ($\approx 10^3$ fm$^3$), we show that in the equilibrium case,
the discontinuity of the order parameter in the first order phase transition region is replaced by smooth crossover, and the corresponding fluctuations are broadened.
Meanwhile, the quartic cumulant $\kappa_4$ of the $\sigma$ field is generally negative near the phase transition line. We further derive the dynamical evolution of the QCD Landau free energy in the Fokker-Plank framework, based on which we deduce the dynamical cumulants of the $\sigma$ field. Assuming the temperature decreases as a known function of time, we numerically evaluate the dynamical cumulants and confirm that the cumulants present clear memory effects. Moreover, the memory effects on the first order phase transition side is stronger than that on the crossover side, and the dynamical cumulants at the hypothetical freeze-out line present rich non-monotonic structures.}

\maketitle

\section {Introduction}

The QCD phase structure has been of intensive theoretical and experimental interest for decades \cite{Cabibbo:1975,Halasz:1998qr,Baym:2002,Fukushima:2011,Aggarwal:2010cw,Nuclear:2008}. In various extreme conditions,  theoretical studies predict different phases of the QCD matter, such as quark-gluon plasma (QGP), hadronic resonance gas (HRG), and color superconductor. Of great interest is the chiral phase transition ($\chi$PT) between the hot QGP phase and the HRG phase \cite{stocker:1986,TDLee:1974,Hatsuda:1985}.
For small baryon chemical potential $\mu$, lattice simulations show that the transition is a broad crossover \cite{Brown:1990,Aoki:2006we,Aoki:2009sc,Karsch:2003jg}. For large $\mu$, the sign problem of lattice QCD prevents full {\it ab initio} simulations. The phenomenological models such as the Nambu-Jona-Lasinio model \cite{Klevansky:1992qe,Fukushima:2004qe,Fu:2007xc,Jiang:2013} and quark-meson model \cite{Schaefer:2007,Schaefer:20072,Herbst:2011}, and non-perturbative methods like Functional Renormalization Group method \cite{Berges:2000ew} and Dyson-Schwinger equation \cite{Roberts:1994dr,Qin:2011} provide relatively complete description on the $\chi$PT, predicting crossover at small $\mu$, first order phase transition at large $\mu$, and the existence of a QCD critical point.

On the other hand, owing to the strong coupling between the chiral $\sigma$ field and quarks in QCD-inspired models, the high-order cumulants of net-proton production are expected to be sensitive to the increase of fluctuations of the $\sigma$ field near the critical point \cite{Stephanov:2008qz,Athanasiou:2010kw}.
The related observables are being measured by the STAR collaboration at the Relativistic Heavy Ion Collider (RHIC) \cite{Adamczyk:2013dal,Luo:2015ewa,STAR:2020tga}. The latest experimental data of $\kappa\sigma^2$ for the net-proton production presents a non-monotonic variation as a function of collision energy for the region $\sqrt{s_{NN}}=7.7-200$ GeV in the Au+Au central collisions \cite{STAR:2020tga}, which partially agrees with theoretical expectations \cite{Stephanov:2008qz}. However, as shown in Ref. \cite{Jiang:2015hri}, by introducing a freeze-out scheme to the hydrodynamics, the model calculations of the equilibrium critical fluctuations of the net protons still failed to qualitatively explain the full experimental data. It presented the limitation of thermal equilibrium assumption and suggested that the critical dynamics should be taken into account for the description of dynamical $\chi$PT in heavy ion collision.

In recent years, to study the dynamics of non-equilibrium fluctuations, including the dynamical evolution of the order parameter field and the diffusion of the conserved charges, different dynamical models have been developed  \cite{Stephanov:2009ra,Mukherjee:2015swa,Jiang:2017mji,Jiang:2017fas,Jiang:2021zla,Sakaida:2017rtj,Nahrgang:2018afz}. The dynamical critical fluctuations consistently present clear memory effects and critical slowing down effects. As a result, both the sign and the magnitude of the high order cumulants can be different from the equilibrium ones, and the kurtosis shows non-monotonic behaviors. To be closer to the experimental process in RHIC, the coupled dynamical evolution of the critical modes and the hydrodynamic background are further developed, like that has been done in the chiral hydrodynamics \cite{Paech:2003fe,Nahrgang:2011mg,Nahrgang:2011vn,Herold:2013bi,Herold:2013qda} and hydro+ \cite{Stephanov:2017ghc,Rajagopal:2019xwg,Du:2020bxp,An:2020vri}. Moreover, other kinds of factors such as the spatially nonuniform temperature (and chemical potential) effects on the fluctuations of the order parameter \cite{Zheng:2021pia}, the non-critical fluctuations \cite{Bluhm:2016byc}, and the proper freeze-out scheme  \cite{Jiang:2015hri,Pradeep:2022mkf} will also influence the theoretical predictions significantly.

As a main component of the critical dynamics, parameterization of the QCD equation of state through the Ising mapping is usually applied to get insight into the universal behaviors in the critical region\,\cite{Mroczek:2020rpm,Monnai:2019hkn,Noronha-Hostler:2019ayj,Parotto:2018pwx}.
However, due to the discontinuities of QCD equation of state on the first order phase transition side,  the critical dynamics remains unclear and requires more exploration. In this paper, we focus on the dynamics of critical fluctuations in the $\chi$PT region (including the first order phase transition and crossover) rather than building the complete dynamical modeling. We provide an alternative way to parameterize the QCD Landau free energy in different phase transition scenarios (which also serves as the basis of our discussion on the spatially nonuniform temperature effects in Ising-like models \cite{Zheng:2021pia}), evaluate the finite size effects of the QCD matter on the order parameter and fluctuations, and develop a distinct set of dynamical equations based on the Fokker-Plank equation to study the dynamical free energy and the dynamical cumulants. The article is structured as following. In Sec.\,\ref{ising}, we linearly parameterize the Landau free energy in the ($\mu$, $T$) plane, where both the crossover and the first order phase transition side are described. In Sec.\,\ref{finit}, we set up the parameters in the free energy and present the results of the equilibrium cumulants in the  system of different volumes. In Sec.\,\ref{dynamical}, we deduce the dynamical free energy based on the Fokker-Plank equation. In Sec.\,\ref{cumulants}, we numerically calculate the non-equilibrium cumulants at different scenarios and on a hypothetical freeze-out line with a fixed volume $V=10^3 \text{ fm}^3$. Finally, in Sec.\,\ref{discussion}, we summarize the main results of this paper and give a discussion.

\section{Parameterization of the free energy}\label{ising}

Because the critical behaviors of the $\sigma$ field  fall into the same universality class as the 3D Ising model \cite{Gavin:1994}, in principle, one can parameterize the QCD equation of state by directly mapping the QCD parameters $(\mu,T)$ to the Ising variables $(r, h)$, as is done in Refs.\,\cite{Guida:1997,Berdnikov:1999ph,Nonaka:2005,Stephanov:2011}. Combined with the dynamical models, there are various interesting dynamical effects presented, like memory effects and universal off-equilibrium scaling behaviors\,\cite{Berdnikov:1999ph,Mukherjee:2015swa,Mukherjee:2016kyu,Nahrgang:2018afz}. However, the early studies focus their discussions of critical dynamics on the crossover side, due to the discontinuities of the equations of state in the first order phase transition region. In this section, we develop an alternative method to directly parameterize the QCD free energy, in which the two phase transition scenarios are unified in a same framework.

According to the Landau theory of phase transition, the free energy in the critical region is supposed to be analytic and obeys the symmetry of the Hamiltonian. Then, the Landau free energy density of the $\chi$PT\,\cite{Pisarski:1984,Gell-Mann:1960} can be generally written in terms of the $\sigma$ field as
\begin{equation}
\Omega[\sigma]=\alpha _{1}(\mu,T)\sigma +\frac{\alpha _{2}(\mu,T)%
}{2}\sigma ^{2}+\frac{\alpha _{3}(\mu,T)}{3}\sigma ^{3}+\frac{\alpha
_{4}(\mu,T)}{4}\sigma ^{4}, \label{pot}
\end{equation}%
which is Taylor expanded as a function of $\sigma$ up to the fourth order. The constant term is omitted because it is irrelevant to the structure of the QCD phase diagram. Note that a zero-momentum mode approximation of the $\sigma$ field is assumed in above, because we focus on the phase transition region in the vicinity of the critical point, where the long-wavelength modes are dominant. The distribution of the critical field is described by the probability distribution function
$P[\sigma] \propto {\exp}\{- \Omega[\sigma] V/T\}$ \cite{Stephanov:2008qz}, where $V$ is the volume of the system. In the chiral limit, $\alpha _{1}=\alpha _{3}=0$, while for the physical world, a finite $\alpha _{1}\sigma $ term is introduced to handle the explicit chiral symmetry breaking of the quark masses \cite{Petropoulos:1999}. The cubic term, $\alpha _{3}\sigma^3$, emerges after the renormalization contributed from the high-momentum modes of the $\sigma$ field. The coefficient of the quartic term, $\alpha _{4}$, is supposed to be positive, in order to sustain the stability of the system.

In the thermodynamic limit ($ V \rightarrow \infty$), the phase structure is fully determined by the global minimum of the free energy \eqref{pot}. It is convenient to present the information of the $\chi$PT by performing a translation transformation for the $\sigma$ field in the free energy, i.e, $\sigma =\tilde{\sigma}+\sigma _{c}$, where
\begin{equation}\label{sigmac}
\sigma_{c}(\mu,T)=-\frac{\alpha _{3}(\mu,T)}{3\alpha _{4}(\mu,T)}.
\end{equation}
Consequently, the cubic term in Eq.\,(\ref{pot}) can be eliminated, and we obtain the translated free energy density (after getting rid of a constant term)
\begin{equation}
\Omega \lbrack \tilde{\sigma}]=\eta _{1}\left( \mu,T \right) \tilde{\sigma}+ \frac{1}{2}\eta _{2}\left( \mu,T \right) \tilde{\sigma}^{2}+\frac{1}{4}\eta
_{4}\left( \mu,T \right) \tilde{\sigma}^{4},  \label{pot2}
\end{equation}%
which has exactly the same form as the free energy density of the Ising model. The redefined coefficients, $\eta _{i}$, can be expressed as functions of the original coefficients $\alpha _{i}$ and the shift $\sigma_c$ by
\begin{eqnarray}
 \eta _{1} &=& \alpha _{1}+\alpha _{2}\sigma _{c}-2\alpha
_{4}\sigma _{c}^{3}, \label{eta1}\\
\eta _{2} &=& \alpha _{2} -3\alpha _{4}\sigma
_{c}^{2},  \label{eta2} \\
\eta _{4} & =& \alpha _{4}.   \label{eta4}
\end{eqnarray}
Now the $\tilde{\sigma}$ field has become an effective order parameter, which behaves just like the spin density in the Ising model.
\begin{figure}[tpb]
\centering
\includegraphics[width=0.9\columnwidth]{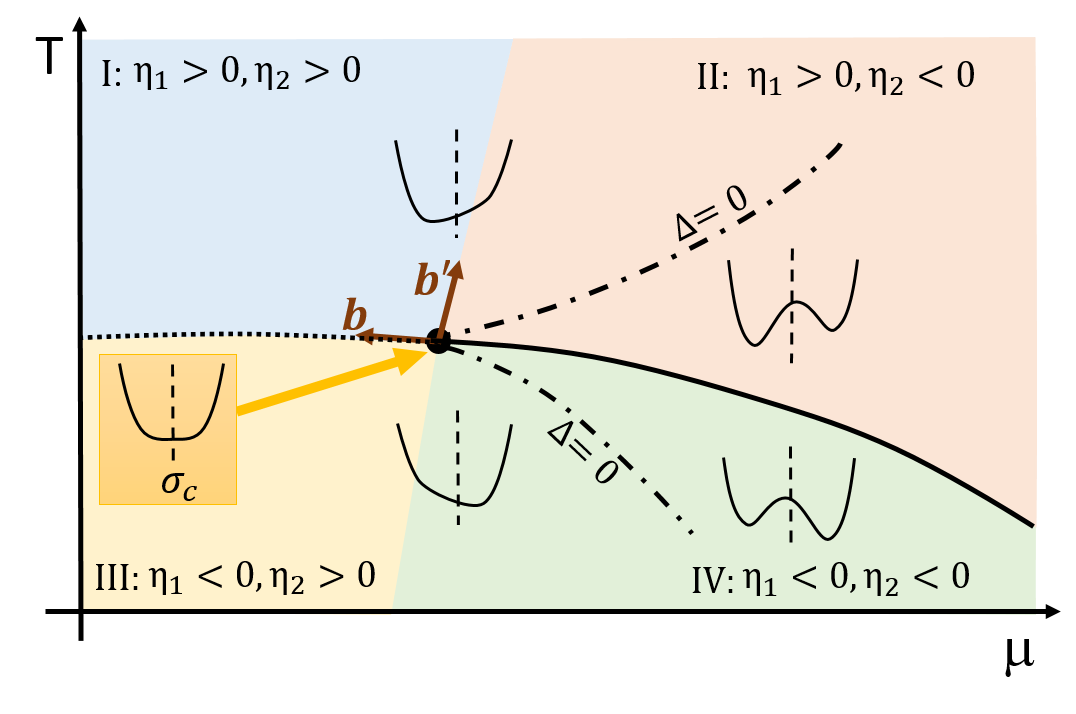}
\caption{A sketch of the QCD phase diagram. The colored 4 zones (I-IV) are sorted according to the sign of $\eta_1$ and $\eta_2$. The dash-dotted lines determined by  $\Delta \equiv (\frac{\eta_1}{2 \eta_4})^2 + (\frac{\eta_2}{3 \eta_4})^3=0$, are the boundary of the region where two local minima coexist in the free energy.}
\label{phase}
\end{figure}

With the translated free energy, we can divide the phase spaces in the $\mu$-$T$ plane into four zones, according barely to the sign of $\eta_{1}$ and $\eta _{2}$. As shown in Fig.\,\ref{phase}, the phase transition line is determined by $\eta_1 =0 $, and the order of the phase transition is determined by the sign of $\eta _{2}$.
At the right part of the diagram (zone II and IV), $\eta _{2}<0$, the phase transition is first order, as there is a coexistence region around the phase transition line, in which two local minima of the free energy coexist. The coexistence region is separated out by the dash-dotted lines, which are determined by $\Delta \equiv (\frac{\eta_1}{2 \eta_4})^2 + (\frac{\eta_2}{3 \eta_4})^3<0$. The physical vacuum locates at the lower minimum of the free energy. This is the typical
feature of the first order phase transition. For the other parts of the $\mu$-$T$ plane where $\Delta>0$, only one minimum exists, and the system undergoes a continuous phase transition (crossover) as $\eta_1$ crosses its zero point. The sign of $\eta _{1}$ determines the location of the global minimum. In general, in the upper regions (zone I and II), with $\eta _{1}>0$, the global minimum is at $\tilde\sigma <0$ (i.e.  $\sigma < \sigma _{c}$); while in the lower regions (zone III and IV), with $\eta _{1}<0,$ the global minimum is at $\tilde\sigma >0$ (i.e.  $\sigma > \sigma _{c}$). At the critical point $(\mu_c,T_c)$,  both $\eta_1$ and $\eta_2$ vanish.

To parameterize the coefficients in $\Omega [\tilde{\sigma}]$ linearly, it is convenient to define two unit vectors $\mathbf{b}=(\cos\theta_b,\sin\theta_b)$ and $\mathbf{b}^{\prime}=(\cos\theta_{b'},\sin\theta_{b'})$ in the $\mu$-$T$ plane. The vector $\mathbf{b}$ is parallel to the tangent of the phase transition curve at the critical point; The vector $\mathbf{b}^{\prime}$ is along the boundary of region I and II in Fig.\,\ref{phase}. Note that $\mathbf{b}$ and $\mathbf{b}^\prime$ are not necessarily orthogonal. The angle between them are determined by the realistic QCD equation of state around the critical point, which is still under study. By projecting the vector $(\mu -\mu _{c},T-T_{c})$ onto the perpendicular vector of $\mathbf{b}$ and that of $\mathbf{b}^\prime$, respectively, the coefficients are linearly parameterized as:
\begin{eqnarray}
\eta _{1}(\mu,T)&=&d_{1}[\left( \mu -\mu _{c}\right)\sin\theta_b
- \left( T-T_{c}\right)\cos\theta_b ] , \label{eta1d1}\\
\eta _{2}(\mu,T)&=&d_{2}[ -\left( \mu
-\mu _{c}\right)\sin\theta_{b'} +\left( T-T_{c}\right)\cos\theta_{b'} ] ,\label{eta2d2}\\
\eta _{4}(\mu,T)&=&d_{4}. \label{eta4d4}
\end{eqnarray}
With these projections, the sign of $\eta _{1}$ and $\eta _{2}$ at different parts (I-IV) can be correctly expressed by simply constraining all the constants to $d_{i}>0$  ($i=1,2,4$), i.e., in the right side of the vector $\bf b$, $\eta_1 >0$, and in the left side of $\bf b'$, $\eta_2 >0$. Note that the magnitude of $d_{i}$ are again determined by the QCD equation of state, which currently remains unknown. In the following, for simplicity and illustration, we treat them as input parameters, and make qualitative calculations and discussions. Along the phase transition curve, $\eta_1(\mu, T) =0$,  the correlation length is $\xi = \eta_2^{-1/2}$ on the crossover side and $\xi = (-2 \eta_2)^{-1/2}$ on the first order phase transition side, which diverges at the critical point where $\eta_2 =0$\,\cite{Fukushima:2011}. In the critical region, we assume that the change of $\sigma_{c}$ on the phase transition line is small and treat the variable $\sigma_{c}(\mu,T)$ as a constant, $\sigma_{c}(\mu,T) = \sigma_{c}(\mu_c,T_c)$, in the zero-order approximation of variables $\mu-\mu_c$ and $T-T_c$. Then, the translated free energy density \eqref{pot2} is fully parameterized by $\mu_c$, $T_c$, $\sigma_{c}$, $d_1$, $d_2$, $d_4$, $\theta_b$, and $\theta_{b'}$. By using the relations Eq.\,(\ref{sigmac}) and Eqs.\,(\ref{eta1})-(\ref{eta4}), the original free energy $\Omega[\sigma]$ (Eq.(\ref{pot})) can also be expressed in the parameterized form.

Through the transformation from Eq.\,\eqref{pot} to Eq.\,\eqref{pot2} and the parameterization Eqs.\,\eqref{eta1d1}-\eqref{eta4d4}, we have built the connection between the QCD free energy and the Ising free energy at the mean-field level. Indeed, Eqs.\,\eqref{eta1d1} and \eqref{eta2d2} describe the dependence of $\eta_1$ and $\eta_2$ on the QCD variables ($\mu$, $T$) in the linear approximation, and the coefficients $\eta_1$ and $\eta_2$ in the translated free energy (Eq.\,\eqref{pot2}) are directly related to the reduced magnetic field and temperature $h$ and $r$ in the Ising model, respectively. Note that the current parameterization of the QCD free energy is only up to the linear order, which works in the region close to the critical point. When treating a larger region in the $\mu$-$T$ plane, higher order expansions of $\mu-\mu_c$, and $T-T_c$ should be taken into account. It is worth stressing that this parameterization can also be applied to the (hadronic) gas-liquid phase transition \cite{Vovchenko:2015pya,Vovchenko:2016rkn} or to other similar scenarios, as long as they belong to the same universality class. With this parametric free energy, we can easily develop approaches to study the equilibrium and non-equilibrium cumulants in different phase transition scenarios, as will be shown in the following sections.

\section{parameter setup and equilibrium cumulants}\label{finit}

In this section, we calculate the equilibrium cumulants based on the above parametric Landau free energy. The cumulants of  the $\sigma$ field  are defined as
\begin{eqnarray}
\kappa _{1}&=&\mu_{1},\\
\kappa _{2}&=&\mu_{2}-{\mu_{1}}^{2},\label{eq:variance}\\
\kappa _{3}&=&\mu_{3}-3\mu_{2}\mu_{1}+2{\mu_{1}}^{3}, \label{eq:variance2} \\ \kappa_{4}&=&\mu_{4}-4\mu_{3}\mu_{1}-3{\mu_{2}}^{2}+12\mu_{2}{\mu_{1}}^{2}-6{\mu_{1}}^{4}, \label{eq:variance3}
\end{eqnarray}
where the moments $\mu_n$ are
\begin{eqnarray}
\mu_n = \langle \sigma^n \rangle =\int d\sigma \sigma^n P[\sigma] {\Big/} \int d\sigma  P[\sigma],
\end{eqnarray}
and the probability distribution function for the $\sigma$ field is
\begin{eqnarray} \label{statis}
P[\sigma] \propto \rm{exp}\{- \Omega[\sigma] V/T\}.
\end{eqnarray}

The numerical calculation of the cumualnts are implemented with the following setup of parameters: $(\mu _{c},T_{c})=(240, 170)$ {MeV}, $\sigma_{c}=50 ~\rm{MeV}$, $d_{1}=3\times 10^{4} ~\rm{MeV}^{2}$, $d_{2}=400 ~\rm{MeV}$, $d_{4}=15$,    $\sin\theta_{b'} = -\cos\theta_b = 0.99$, and $\cos\theta_{b'} = \sin\theta_b= 0.141$. Here, $\mathbf{b} \perp \mathbf{b}^\prime$ is set for simplicity. The choice of these parameters are constrained by comparing with the effective potential and cumulants from the linear sigma model with constituent quarks in \cite{Paech:2003fe}, with which the equilibrium values of the cumulants in the phase transition region are approximately of the order $\kappa_n \sim 10^n~{\rm MeV}^n$.

Note that the discussion and parameterizations of the free energy density (\ref{pot}) in above is in the thermodynamic limit.  In reality, especially in the experimental environment, the size of the QGP fireball in the heavy ion collision is finite (the typical volume of the fireball is approximately $V=10^3~ {\rm fm}^3$). The finite size has a direct influence on the renormalization process. Consequently, the coefficients $\alpha_i$ in Eq.\,(\ref{pot}) depend on the volume, and further, the parameters $\mu_c$, $T_c$, $\sigma_c$, $d_1$,  $d_2$, $d_4$, $\theta_b$, and $\theta_{b'}$ are also volume-dependent. There is a standard process for the renormalization of the coefficients of the free energy\,\cite{brezin1985npb}. However, even through the volume will change the magnitude of the parameters, the parameterized form of the Landau free energy density will not be changed. Therefore, in this article, we will not discuss the finite size corrections to the free energy density itself but simply utilize the same parameter set for different volumes to study the behaviors of the cumulants.

\begin{figure}[tpb]
\centering
\includegraphics[width=\columnwidth]{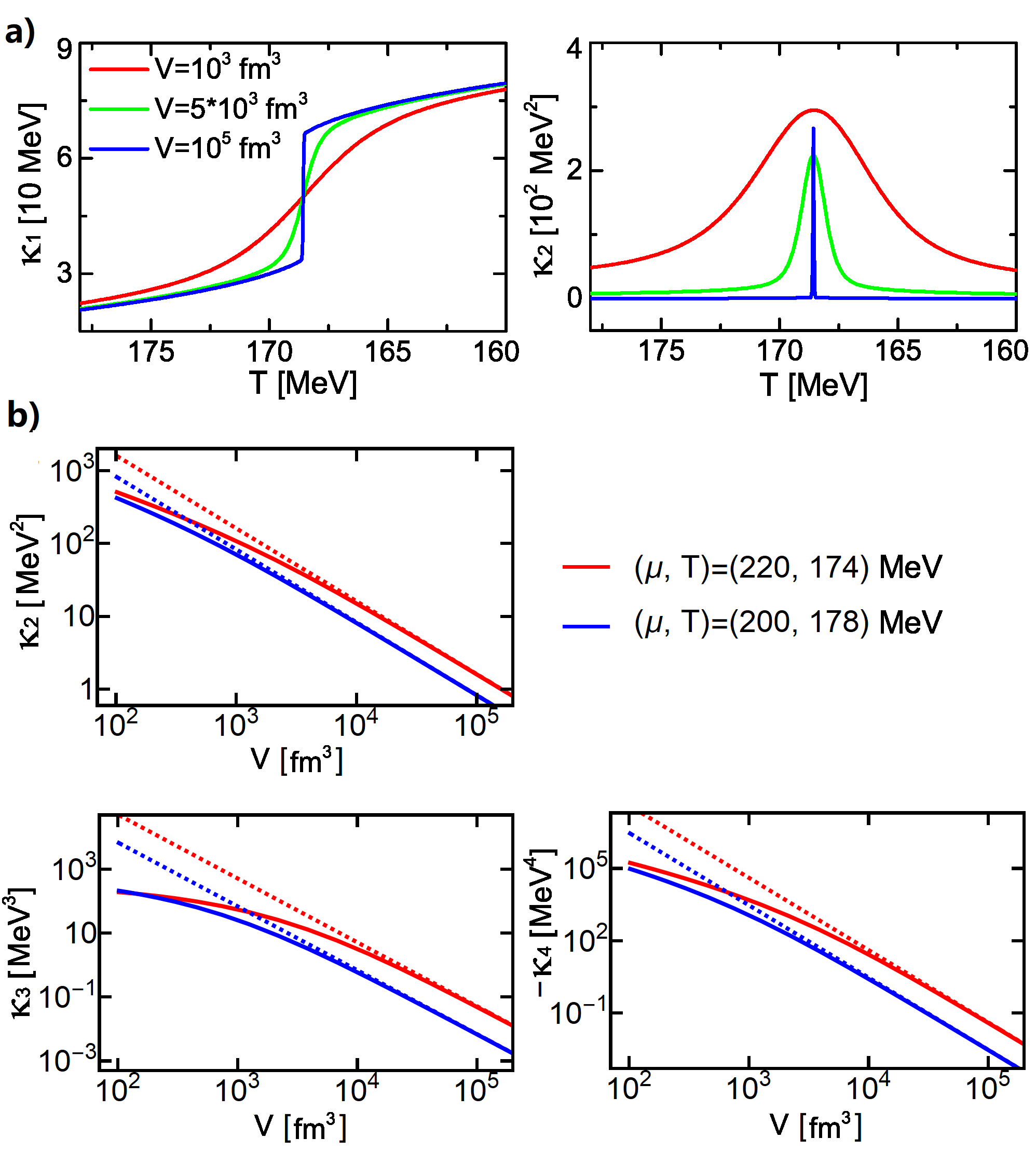}
\caption{(a) The cumulants $\kappa_1$ and $\kappa_2$ of the $\sigma$ field as functions of the temperature with different fixed volumes  in the first order phase transition regime. The chemical potential is fixed at $\mu= 250  ~\rm{MeV}$. (b) The $\kappa_2, \kappa_3$ and $\kappa_4$ as functions of volume in the crossover regime. The solid lines are given by Eq.\eqref{eq:variance}-\eqref{eq:variance3}. The dotted lines represent fluctuations determined by the Gaussian approximation, $\tilde{\kappa_2} = T\xi^2/V$, and the perturbative results $\tilde{\kappa_3}$ and $\tilde{\kappa_4}$ \cite{Mukherjee:2015swa}.}
\label{vol-effect}
\end{figure}

Even without considering the influence on the free energy density from the renormalization of size, we still have to face the finite-volume effects on the fluctuations of the $\sigma$ field as shown in the probability function \eqref{statis}. In Fig.\,\ref{vol-effect}(a), we present the first two order cumulants of the $\sigma$ field  with respect to temperature for different volumes. The chemical potential is fixed at $\mu=250$ MeV so the system undergoes a first order phase transition as the change of temperature. For a large enough volume (say $10^5 \rm{fm}^3$), the expectation value $\kappa _{1}=\langle \sigma \rangle$ is discontinuous because the barrier of the free energy $(\Delta\Omega V)$ can not be overcome by the thermodynamic fluctuations, i.e., $\Delta\Omega V \gg T$.  The $\sigma$ field locates only near the global minimum point and the variance $\kappa_2$ is also suppressed by the volume.
Specifically, on the phase transition line $\eta_1 = 0$,  the correlation length of the $\sigma$ field is $\xi =  (-2 \eta_2)^{-1/2} $ and the barrier is $\Delta\Omega V = \eta^2_2 V/4\eta_4$. Thus, the discontinuity of the first order phase transition becomes evident when $V \gg 16 T \eta_4 \xi^4$. For a smaller system volume, the value of $T/V$ in the probability distribution function $P[\sigma]$ grows.  As a result, in the first order phase transition region, the $\sigma$ field is allowed to be at both of the local minimum points of the free energy with considerable probability, which leads to the enhancements of variance $\kappa_2$. Meanwhile, the expectation value $\kappa _{1}$ deviates from the global minimum of the free energy, and its discontinuity is rounded. Similar finite-size rounding effects of a first order phase transition can also be found in Refs.\cite{Imry1980prb,Spieles:1997ab,Zabrodin:1998vt}. This hints that for a small system, the perturbation theory around the global minimum does not work well anymore, and the equation of states obtained in the thermodynamic limit \cite{Stephanov:2008qz,Mukherjee:2015swa} are not suitable for the system with small size.

In Fig.\,\ref{vol-effect}(b), we also plot $\kappa_2$-$\kappa_4$ as functions of the volume of the system for $(T,\mu)$ locating in the crossover regime, and make comparisons with the Gaussian fluctuations $\tilde{\kappa_2} = T\xi^2/V$ and the perturbative non-Gaussian cumulants $\tilde\kappa_3=-2\lambda_3 \left({T}/{V}\right)^2\xi^6$ and $\tilde\kappa_4=6 \left({T}/{V}\right)^3[2(\lambda_3 \xi)^2 -\lambda_4]\xi^8$ \cite{Mukherjee:2015swa}. For large volume, the variance $\kappa_2$ is consistent with  $\tilde{\kappa_2}$. But as the decreasing of the volume, the non-Gaussian corrections are enhanced as the widening of the variance, and lead to a significant deviation of $\kappa_2$ from $\tilde{\kappa_2}$. The Gaussian approximation is valid when the non-Gaussian contributions are small. Expanding Eq.\,\eqref{pot2} around the extreme point $\tilde\sigma_m$ where $\partial_{\tilde\sigma}\Omega = 0$, the third and forth order coupling constants of the fluctuation $\delta\tilde\sigma$ defined as in Refs.\,\cite{Stephanov:2008qz,Mukherjee:2015swa} become $\lambda_3= 3 \eta_4 \tilde\sigma_m $
and $\lambda_4 = \eta_4$. The corresponding conditions for the validity of the Gaussian approximation are $\lambda_3 ({\delta\tilde\sigma})^3 V/3T \ll 1$ and  $\lambda_4 ({\delta\tilde\sigma})^4 V/4T \ll 1$. The typical magnitude of $\delta\tilde\sigma$ is of the order of $\sqrt{\tilde{\kappa_2}}$. Thus, the conditions become
$V \gg (\eta_4 \tilde\sigma_m)^2  \xi^6 T$ and  $V \gg \eta_4 \xi^4 T/4$. Specially on the phase transition line $\eta_1 =0$, we have $\tilde\sigma_m =0$ and the first condition is satisfied automatically. In the condition of large volume, the higher order cumulants $\kappa_3$ and $\kappa_4$ also become consistent with the perturbative results as shown in the figure.

\begin{figure}[tbp]
\centering
\includegraphics[width=\columnwidth]{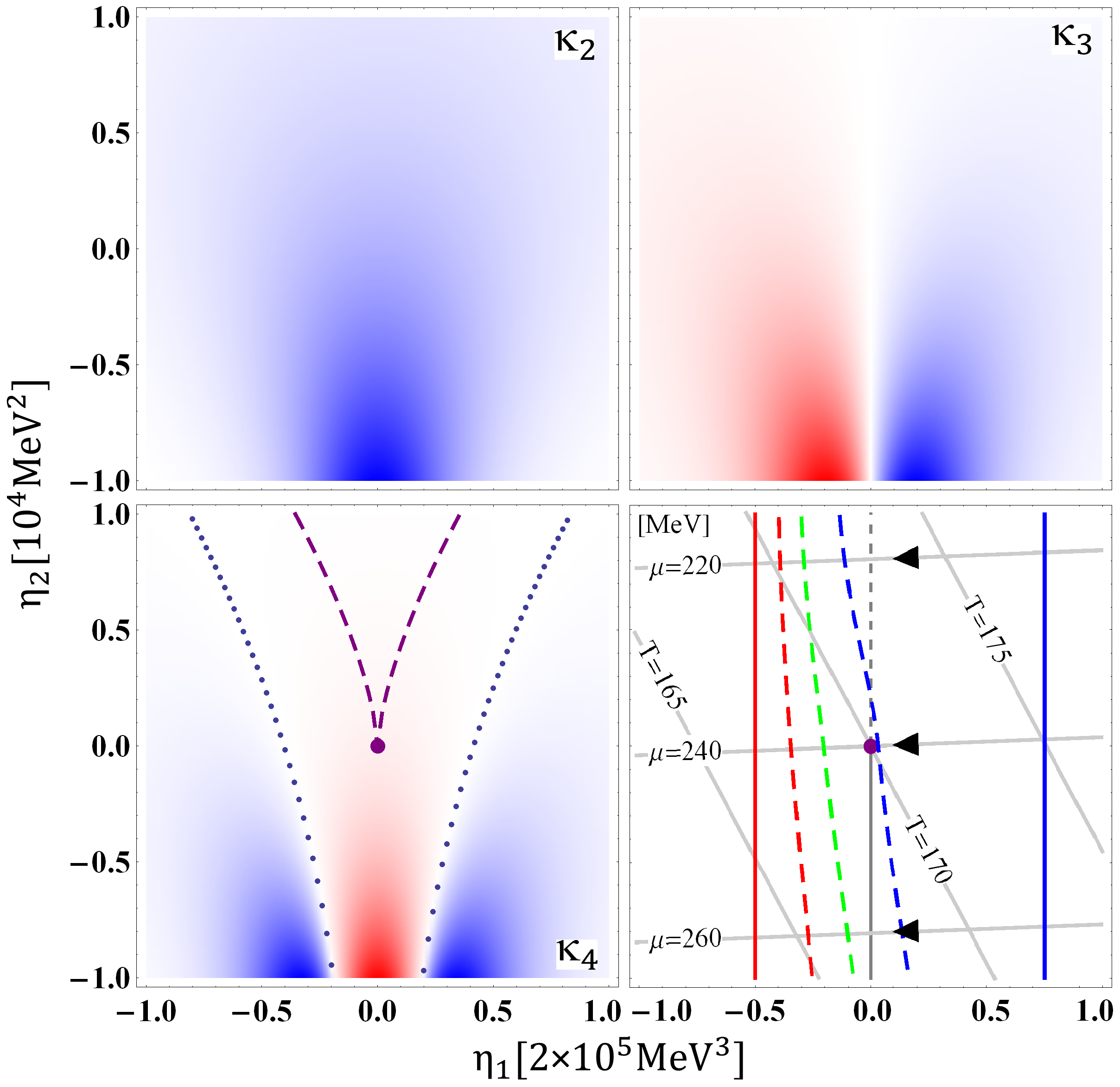}
\caption{The equilibrium cumulants $\kappa_2 - \kappa_4$ in the $\eta_1$-$\eta_2$ plane for $V=10^3~ \rm{fm}^3$. The colors red and blue refer to negative and positive sign of the cumulants, respectively, and darker colors correspond to larger magnitudes of the cumulants. $\eta_2<0$ corresponds to the first order phase transition region and $\eta_2>0$ corresponds to the crossover region. $\eta_1=0$ refers to the phase transition line. In the subfigure $\kappa_4$, the dotted blue lines mark the boundary where $\kappa_4$ changes the sign; while as a comparison, the dashed purple lines are the boundary in the thermodynamic limit case.
The bottom-right subfigure shows the setup of the dynamical evolution in Sec.\,\ref{cumulants}, and visualize the memory effects with different relaxation rates. The grey lines are the contours of $\mu$ and $T$ in the $\eta_1$-$\eta_2$ plane. The dynamical evolution are set to be along fixed chemical potentials, while the blue and red solid line are the starting and hypothetical freeze-out line for the dynamical evolution. The evolution direction is marked by arrows. The dashed red (green, blue) line, on which the equilibrium $\kappa_1$ equals to the dynamical $\kappa_1$ on the solid red line (hypothetical freeze-out line) with relaxation rate $\tau_{\rm{rel}} =0.05 (0.1,0.2 ~\rm{fm})$, represents the effective freeze-out line due to the memory effects of dynamics.
}
\label{cumus}
\end{figure}

In the following calculations, we fix the system volume to the typical size of the QGP fireball, $V=10^3~ \rm{fm}^3$. In Fig.\,\ref{cumus}, we present the density plots of the equilibrium cumulants $\kappa_2 - \kappa_4$ in the $\eta_1$ - $\eta_2$ plane.
The magnitude of $\kappa_2, \kappa_3$, and $\kappa_4$ in the first order phase transition region ($\eta_2<0$) are generally larger than that in the crossover region ($\eta_2>0$) for a given small $\eta_1$. Similar to Ref.\,\cite{Stephanov:2008qz}, $\kappa_3$ changes its sign after crossing the phase transition line. On the other hand, unlike the results in Ref.\,\cite{Stephanov:2008qz}, $\kappa_4$ is generally negative (red color region)
near the phase transition line ($\eta_1 = 0$), for both the crossover side and the first order phase transition side. The change of sign for $\kappa_4$ happens at the  dotted blue lines, inside which the red color region is much larger than the region rounded up by the dashed purple lines (refers to the sign-change line of $\kappa_4$ in the thermodynamic limit). As explained in the above paragraphs, the broadening of the negative $\kappa_4$ region on the first order phase transition side is due to the rounding effects, since the two-peak shape of $P[\sigma]$ (for the double-well of $\Omega[\sigma]$ with small volume) has less kurtosis than the Gaussian distribution\,\cite{Stephanov:2011}. While in the thermodynamic limit for the first order phase transition region, even through the free energy  $\Omega[\sigma]$ also has the double-well shape, one peak of $P[\sigma]$ is strongly enhanced by the large volume comparing to the other one, thus $P[\sigma]$ shapes like a one-peak distribution and only the region near the global minimum of $\Omega[\sigma]$ is distributed. As a result,  $\kappa_4$ is negative only in the crossover region in the thermodynamic limit \cite{Stephanov:2011}, striking contrast to the results in systems of small size.

\section{Dynamical equations for the free energy}\label{dynamical}

In this section, we employ the Fokker-Plank equation to study the dynamical behaviors of the parameterized QCD free energy. The Fokker-Plank equation for the dynamical probability distribution (denoted by $P[\sigma;t]$) of the $\sigma$ field is\,\cite{Mukherjee:2015swa}
\begin{equation}\label{evolution}
\partial_t P[\sigma;t] = - \frac{1}{m_\sigma^2 \tau_{\rm{eff}}}
\partial_\sigma \left\{\partial_\sigma\left(\Omega[\sigma;t] -
\Omega_0[\sigma] \right) P[\sigma;t] \right\},
\end{equation}
where $\Omega[\sigma;t]\equiv -\frac{T}{V} {\rm ln} P[\sigma;t]$ is the dynamical free energy density and $\Omega_0 \left[\sigma \right]$ is the equilibrium free energy density at the point $(\mu(t), T(t))$ (see Eq.\,(\ref{pot})). The parameter $\tau_{\rm{eff}}$ is the effective relaxation rate. $m_\sigma$ is the equilibrium mass of the $\sigma$ field, which is defined as
\begin{equation}
    m_\sigma^2 =\frac{d^2 \Omega_0[\sigma]}{d\sigma^2}{\big |}_{\sigma =\sigma_0},
\end{equation}
where $\sigma_0$ is the global minimum of $\Omega_0[\sigma]$. The following calculation assumes that the dependence of the relaxation rate $\tau_{\rm{eff}}$ on the equilibrium correlation length $\xi_{\rm{eq}}(\mu, T)$  satisfies
\begin{equation}
    \tau_{\rm{eff}} = \tau_{\rm{rel}}\left(\frac{\xi_{\rm{eq}}}{\xi_{\rm{ini}}}\right)^z,
\end{equation}
where $\xi_{\rm{eq}} = m_\sigma^{-1}$ \cite{Mukherjee:2015swa}. Here
$\tau_{\rm{rel}}$ and $\xi_{\rm{ini}}$ are the initial  relaxation rate and the initial equilibrium correlation length, respectively. The value $z=3$ is given by the dynamical critical exponent of Model H \cite{Hohenberg:1977,Son:2004iv}.

The time evolution equation of $\Omega[\sigma;t]$ is deduced from Eq.\,\eqref{evolution}, and we obtain
\begin{eqnarray}
\partial_t \Omega[\sigma;t] &=& \Omega[\sigma;t] \partial_t\left(%
\rm{ln}\frac{T}{V}\right)+ \frac{T}{V} \frac{\partial_\sigma^2 \left(\Omega[\sigma;t]- \Omega_0[\sigma] \right) }{m_\sigma^2
\tau_{\rm{eff}}}   \nonumber \\
&& - \frac{ \partial_\sigma\left(\Omega[\sigma;t]- \Omega_0%
[\sigma ] \right)\partial_\sigma \Omega[\sigma;t] }{m_\sigma^2
\tau_{\rm{eff}}}.  \label{FP}
\end{eqnarray}
By denoting $\Omega[\sigma;t]= \sum_i \alpha_i(t)\sigma^i /i$, the time evolution of the coefficient $\alpha_i(t)$ becomes,
\begin{eqnarray} \frac{d\alpha_i(t)}{d t} &= & \alpha_i(t)
\partial_t\left(\rm{ln}\frac{T}{V}\right)+ \frac{i (i+1) T}{V} \frac{[
\alpha_{i+2}(t)- \alpha_{i+2}^0]}{m_\sigma^2 \tau_{\rm{eff}}}  \nonumber\\
&& - \sum_{j=0}^{i} \frac{i \alpha_{i-j+1}(t) [\alpha_{j+1}(t)- \alpha_{j+1}^0
]}{m_\sigma^2 \tau_{\rm{eff}}},  \label{FP2}
\end{eqnarray}
where $\alpha_{j}^0 (j=1,\cdots,4)$ are the coefficients for the equilibrium free energy density $\Omega_0[\sigma]$. Note that different coefficients of the $\sigma$ field couple with each other in the time-evolution equations. As a result, the terms of $\sigma^i$ with power $i> 4$ emerge automatically after evolution, even though they vanish at the initial setting. As before, by assuming the contributions from higher power terms are negligible, the free energy is cut off till the fourth order. By setting up the parameters of the dynamical system, we can numerically solve the coupled equations in above, and study the dynamical behaviors of the physical quantities like the cumulants.

\section{ non-equilibrium cumulants}\label{cumulants}

\begin{figure}[tpb]
\centering
\includegraphics[width=\columnwidth]{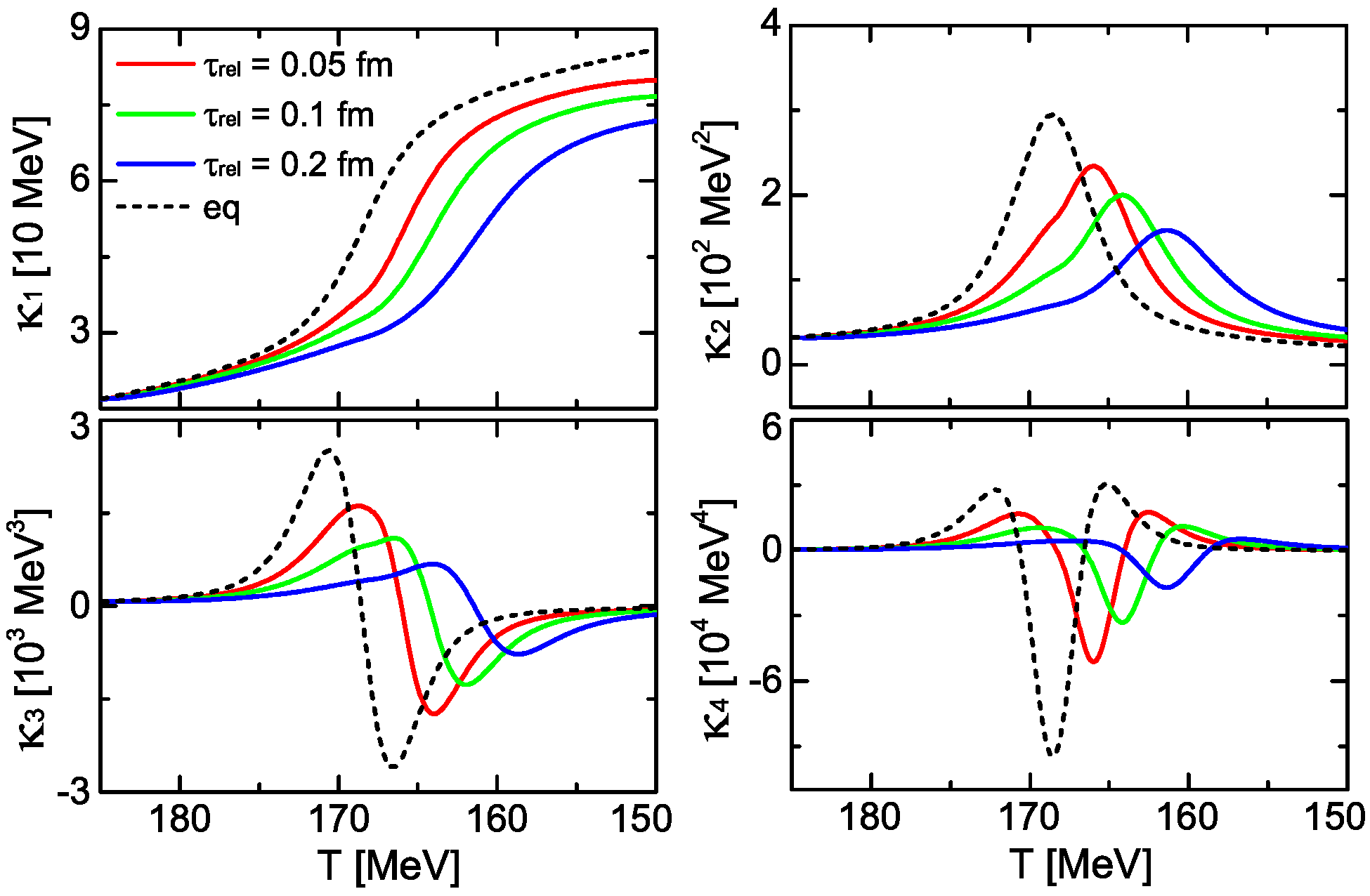}
\caption{The non-equilibrium cumulants with respect to the temperature, for different relaxation rates. Here $\mu=250 ~\rm{MeV}$ and $V = 10^3 ~\rm{fm}^3$. }
\label{dyn-effect}
\end{figure}

As we have obtained the time evolution equation \eqref{FP2} of the QCD free energy, we can study the corresponding dynamical cumulants. For simplicity, in the following calculations, we suppose that both the chemical potential and the volume ($V = 10^3 ~\rm{fm}^3$) are fixed during the time evolution\footnote{With our current setup of the parameters, the time evolution in the critical region is very fast. Hence, the change of the volume during the expansion is neglected. In realistic fireball system, the changing of volume must be included.}.
The temperature is assumed to decrease as a function of time:
\begin{equation}
    T(t)=T_{\rm{ini}} \left( \frac{t + t _{\rm{ref}}}{t _{\rm{ref}}}\right)^{-\lambda},
\end{equation}
where $T_{\rm{ini}}$ is the initial temperature, $t _{\rm{ref}} = 10~\rm{fm}$ is a reference time and the exponent is set to be $\lambda = 0.45$ \cite{Mukherjee:2015swa}.

In Fig.\,\ref{dyn-effect}, we plot the evolution of the cumulants for different relaxation rates, starting from an initial temperature $T_{\rm{ini}}=185 ~\rm{MeV}$.
The monotonicity of the equilibrium cumulants are exactly memorized by the non-equilibrium cumulants, and reproduced in the later evolution process. As the increase of relaxation rate, the peaks and dips of the non-equilibrium cumulants are suppressed, and the typical structure for each order of the cumulants appears in a later time.
Note that in the current finite size system, the memory effects shown in Fig.\,\ref{dyn-effect} is similar for different chemical potentials for both the crossover and the first order phase transition cases.

\begin{figure}[tpb]
\centering
\includegraphics[width=\columnwidth]{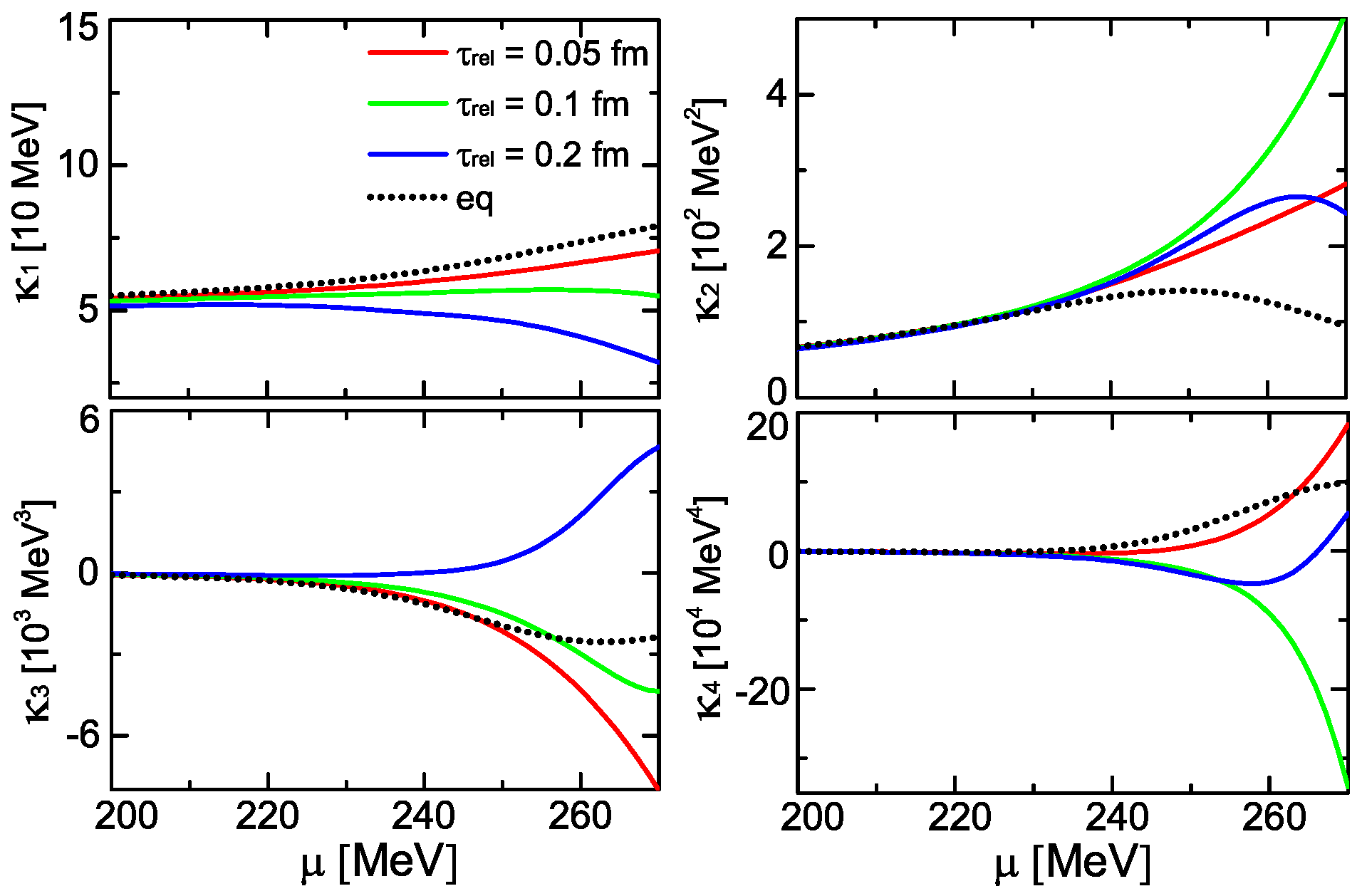}
\caption{The non-equilibrium cumulants with respect to the chemical potential on the freeze-out line, for different relaxation rates. }
\label{dyn-effect2}
\end{figure}

Next, we study the dynamical result of the $\sigma$'s cumulants at a hypothetical freeze-out line. The initial temperature and the hypothetical freeze-out temperature are supposed to be about $5.05~\rm{MeV}$ above and $3.37~\rm{MeV}$ below the phase transition temperature, respectively (i.e. $\eta_1=1.5\times10^5 ~\rm{MeV}^3$ for the initial state and $\eta_1=-1.0\times10^5 ~\rm{MeV}^3$ for the final state, marked as blue and red lines in the bottom-right subfigure in Fig.\,\ref{cumus}).
In Fig.\,\ref{dyn-effect2} we present the non-equilibrium cumulants with respect to the chemical potential. With different relaxation rates, the non-equilibrium cumulants at large chemical potentials (mainly on the first order side) significantly deviate from the equilibrium cumulants, and develop rich structures. Especially, for the blue line with $\tau_{\rm{rel}} = 0.2~\rm{fm}$, $\kappa_3$ change its sign, and $\kappa_4$ presents non-monotonic behaviors compared to the equilibrium ones.
These rich structures are suggestive for the understanding of the STAR data\,\cite{Luo:2015ewa}.

The dynamical behaviors of the cumulants can be understood  from the equilibrium cumulants by considering the memory effects. Let us define an effective freeze-out temperature (for each chemical potential) where the equilibrium expectation of the $\sigma$ field, $\kappa_1$, equals to the dynamical $\kappa_1$ on the freeze-out point. The effective freeze-out lines for different relaxation rates are shown in the bottom-right subfigure of Fig.\,\ref{cumus} (The dashed red, green and blue lines). From the equilibrium cumulants $\kappa_2$-$\kappa_4$ on the effective freeze-out lines,  we can approximately read the sign of the non-equilibrium cumulants in Fig.\,\ref{dyn-effect2}. Specifically, for $\kappa_3$, when the relaxation rate is large, say $\tau_{\rm{rel}} =0.2 ~\rm{fm}$, the effective freeze-out line in the crossover region is basically at the left side of the phase transition line ($\eta_1 <0$), while in the first order phase transition region, it is still at the right side of the phase transition line ($\eta_1 >0$), leading to a positive dynamical $\kappa_3$ (see $\kappa_3$ in Fig.\,\ref{cumus}). For $\kappa_4$, when $\tau_{\rm{rel}} =0.1 ~\rm{fm}$, the effective freeze-out line falls into the negative $\kappa_4$ region; when $\tau_{\rm{rel}} =0.2 ~\rm{fm}$ ($\tau_{\rm{rel}} =0.05 ~\rm{fm}$), the effective freeze-out line crosses the right sign-change line of the $\kappa_4$ in $\eta_1$-$\eta_2$ plane (see $\kappa_4$ in Fig.\,\ref{cumus}).  In general, the memory effects in the first order phase transition region is more significant than that in the crossover region (i.e. the cumulants record earlier information at larger $\mu$), which is because in the first order phase transition region, the barrier in the free energy strongly delays the dynamical evolution of the $\sigma$ field.

\section{Discussion}\label{discussion}

In summary, we parameterize the Landau free energy of the $\chi$PT by mapping it to the Ising free energy.  With the parametric free energy, we study the equilibrium cumulants in a finite size system. The volume of the system significantly influences the fluctuations of the $\sigma$ field. In the equilibrium case, we find that for a typical QGP fireball size with volume $V =10^3$ fm$^3$, the probability distribution of the $\sigma$ field is broadened near the phase transition line, leading to enhancements of the fluctuations and rounding of the discontinuities on the first order phase transition side. Moreover, the fourth order cumulant $\kappa_4$ of the $\sigma$ field is universally negative in the phase transition region on both the crossover and the first order phase transition side. Compared to the crossover region, all cumulants in the first order phase transition region are enhanced, due to the two-peak shape of $\sigma$'s distribution in the finite system (where the barrier of the free energy density is of the order of $T/V$). 
Utilizing the Fokker-Plank equation, we derive the real-time evolution of the parametric free energy, and further, by setting the cooling down process of the system, we analyze the time evolution of non-equilibrium cumulants at different phase transition scenarios and non-equilibrium cumulants on the hypothetical freeze-out line. We find that earlier information about the cumulants are recorded in the first order phase transition region, compared to the crossover region. Like the earlier studies, the dynamical cumulants can be different from the equilibrium ones from both the magnitude and the sign, and can be nonmonotonic on the hypothetical freeze-out line.

Note that in our study, we adopt the zero mode approximation which assumes  the order parameter is uniform in space. The nonzero modes can be further taken into account if one employs the Ginzburg-Landau free energy. Moreover, due to the flexibility of our parameterization, the parametric free energy and its dynamical evolution can be easily combined with a realistic dynamical model (like that in chiral  hydrodynamics \cite{Nahrgang:2011mg,Nahrgang:2011vn,Herold:2013bi,Herold:2013qda} or Hydro+ \cite{Stephanov:2017ghc}), to study the full dynamical process with phase transition.

Now compare our method with the former studies on the non-equilibrium cumulants in Refs.\,\cite{Mukherjee:2015swa,Jiang:2017mji,Jiang:2017fas,Jiang:2021zla}. In Ref.\,\cite{Mukherjee:2015swa}, they presented pioneering studies on the real-time evolution of non-Gaussian cumulants derived from the Fokker-Plank equation. The QCD equation of state near the critical point in the thermodynamic limit is inspired from the Ising model and the parameterization is realized by linearly mapping QCD variables $(\mu,T)$ to the Ising variables $(r, h)$. The resulted equation of state and the corresponding equilibrium cumulants could contain the quantum correction from the fluctuations of the order parameter. However, the volume effects on the equation of state and cumulants are fully omitted, which is valid when the correlation length is significantly smaller than the size of the system. In this case, the discontinuities of the equilibrium distribution function of the $\sigma$ field hinder the simulation of critical dynamics with the Fokker-Plank equation in the first order phase transition region, which assumes the slight deviations of dynamical probability distribution from the equilibrium one. As shown in our study, when the size is small enough, the finite size effects becomes significant especially in the first order phase transition region. The rounding effects justifies the application of the Fokker-Plank equation in the first order phase transition region.
In Refs.\,\cite{Jiang:2017mji,Jiang:2017fas,Jiang:2021zla}, the dynamical evolution of cumulants on different phase transition scenarios were studied based on event-by-event simulations of Langevin equation. The effective potential of $\sigma$ is obtained by evaluating the linear sigma model. The evolution takes account of the fluctuation effects (long wavelength modes) in the real space. However, the shortcoming of this approach is the location of the critical point is fixed by the model parameters of the linear sigma model,  which is even far below the chemical freeze-out line determined by the statistical model and experimental data. Thus the dynamical equations of the critical fluctuations in Ref.\,\cite{Jiang:2017mji,Jiang:2017fas} could not be combined with the hydrodynamic equations, and describe the realistic experimental process.
In recent studies, the parameterization of QCD equation of state with nonlinear corrections (consistent with the lattice results)
have been extended from the critical region to a broader region in the $\mu$-$T$ plane \cite{Monnai:2019hkn,Noronha-Hostler:2019ayj,Parotto:2018pwx}. Similar further extension of the parameterization of the free energy density will be important for studying the dynamic evolution of critical fluctuations at RHIC.

\section*{Acknowledgments}
Jiang thanks Volodymyr Vovchenko, and Huichao Song for discussion and comments, and Frankfurt Institue for Advanced Studies at Frankfurt am Main for hospitality, where part of this work has been done. Jiang acknowledges the support from the NSFC under grant no. 12105223, St\"ocker acknowledges the support from the Walter Greiner Gesellschaft zur F\"orderung der physikalischen Grundlagenforschung e.V. through the Judah M. Eisenberg Laureatus Chair at Goethe Universit\"at, and Zheng acknowledges the support from the NSFC under grant no. 12175180 and the research start-up funding from Northwest University.

\end{document}